\documentclass[aps,prd,preprint,showkeys,nofootinbib,amsmath,amssymb]{revtex4}
\usepackage{epsf}
\usepackage{color}
\usepackage[rgb,dvipsnames]{xcolor}
\usepackage{dcolumn}
\usepackage{bm}
\everymath{\displaystyle}

\begin{document} 

\title{Zitterbewegung of massless particles}

\author{\firstname{Alexander J.}~\surname{Silenko}}
\email{alsilenko@mail.ru} \affiliation{Bogoliubov Laboratory of Theoretical Physics, Joint Institute for Nuclear Research,
Dubna 141980, Russia,\\Institute of Modern Physics, Chinese Academy of
Sciences, Lanzhou 730000, China,\\Research Institute for
Nuclear Problems, Belarusian State University, Minsk 220030, Belarus}

\begin {abstract}
Zitterbewegung of massless particles with an arbitrary spin is analyzed in various representations. Dynamics of the group velocity of a massless particle as a whole and of the corresponding radius vector is determined. This radius vector defines any fixed point of the envelope of the moving wave packet characterizing the particle and its group velocity differs from the group velocities of any points of the wavefront. We consider free massless scalar and Dirac particles, the photon, and massive and massless particles with an arbitrary spin and describe them in different representations. For particles with an arbitrary spin, the generalized Feshbach-Villars representation and the Foldy-Wouthuysen one are used.
Zitterbewegung takes place in any representation except for the Foldy-Wouthuysen one. Formulas describing the motion of a ``trembling'' free particle are the same in any representation. In the Foldy-Wouthuysen representation, the operators of the velocity and momentum of a free particle are proportional and Zitterbewegung does not take place. Since the radius vector (position) and velocity operators are the quantum-mechanical counterparts of the classical position and velocity just in the Foldy-Wouthuysen representation, Zitterbewegung is not observable. The same conclusion has been previously made for free massive particles. For relativistic massive particles with spins 0, 1/2, 1 and massless particles with spins 0, 1/2 in arbitrarily strong electromagnetic fields, independent of the external fields Zitterbewegung does not appear in the Foldy-Wouthuysen representation either. This conclusion is made for leading terms in the Hamiltonian proportional to the
zero and first powers of the Planck constant and for such terms proportional to $\hbar^2$ which describe contact interactions. 
\end{abstract}


\maketitle

\section{Introduction}

Zitterbewegung is one of the most important and widely discussed problems of quantum mechanics (QM).
It is a well-known effect consisting in a superfast trembling motion of a free particle. This effect has been first described by Schr\"{o}dinger \cite{ZitterbewegungSc} for relativistic electrons in a free space as a result of an interference between positive and negative energy states. This effect is also known for a scalar particle \cite{ZitterbewegungGG,ZitterbewegungFF,ZitterbewegungPPNL} and for a massive Proca one \cite{ZitterbewegungPPNL,ZitterbewegungJPhysConf}. There are plenty of works devoted to Zitterbewegung. In the past, Schr\"{o}dinger's interpretation of Zitterbewegung was generally accepted \cite{ZitterbewegungBD,Messiah,Thaller}. Evidently, Zitterbewegung appears due to mixing states with positive and negative total energies. The Foldy-Wouthuysen (FW) transformation eliminates this mixing and, therefore, Zitterbewegung. Since just operators and equations of motion in the FW representation are counterparts of corresponding classical variables and equations (see Ref. \cite{PRAFW} and references therein), Zitterbewegung is not an observable physical effect. We focus our attention on papers presenting a correct analysis of the observability of this effect for Dirac and scalar particles. The correct conclusions
about the origin and observability of this effect have been made in Refs. \cite{ZitterbewegungFF,ZitterbewegungPPNL,ZitterbewegungJPhysConf,ZitterbewegungKrekora,Zitterbewegungbook,ZitterbewegungOC,PRAFW,ZitterbewegungDeriglazov,ZitterbewegungKobakhidze}. However, Zitterbewegung of massless particles has not been thoroughly studied. A probable cause is a demonstration by Newton and Wigner \cite{NewtonWigner} and Wightman \cite{Wightman} for the photon that it cannot be strictly localized according to natural criteria. However, the group velocity of massless particles and the corresponding radius vector can be determined and a rigorous consideration of their Zitterbewegung can be fulfilled. Such a consideration is a goal of the present study.
It has been concluded in some precedent investigations that Zitterbewegung of free photons exists \cite{Unal,Kobe}. The existence of this effect is denied in Refs. \cite{ZhiYongWang1,ZhiYongWang2,ZhiYongWang3} but it is noted that Zitterbewegung exists and can in principle be observed for interacting photons. These conclusions have been based on the heuristic forms of the Dirac-like consideration \cite{Unal,Kobe} and on the Dirac-like quantum-mechanical equation (without the FW transformation) \cite{ZhiYongWang2}. The possibilities of a rigorous quantum-mechanical determination of the appropriate position operator and the FW transformation of the Dirac-like quantum-mechanical equation have not been realized in precedent investigations of Zitterbewegung of the photon. 

The paper is organized as follows. In the next section, we explain previously obtained results for Zitterbewegung of massive fermions and bosons. Distinguishing features of a quantum-mechanical description of massless particles are considered in Sec. \ref{masslessQM}. In Sec. \ref{ScalarD}, we study Zitterbewegung of massless scalar and Dirac particles. Quantum mechanics of the free photon is presented in Sec. \ref{QMphotn} and Sec. \ref{photons} is devoted to the analysis of its Zitterbewegung. In Sec. \ref{arbitrary}, we consider Zitterbewegung of particles with an arbitrary spin in the generalized Feshbach-Villars (GFV) representation. The problem of Zitterbewegung in external fields is analyzed in Sec. \ref{External}. Finally, we summarize the obtained results in Sec. \ref{Summary}.

The system of units $\hbar=1, c=1$ and the standard denotations of the Dirac matrices (see, e.g., Ref. \cite{BLP}) are used. Hereinafter, $\bm p\equiv-i\hbar\nabla$ denotes the momentum operator.

\section{Previously obtained results for massive fermions and bosons}\label{Previous}

In our short review of previously obtained results, we follows Ref. \cite{ZitterbewegungPPNL}. The Dirac Hamiltonian for a free spin-1/2 particle is given by
\begin{equation}
{\cal H}_{D}=\beta m+\bm\alpha\cdot\bm p
\label{HamD}
\end{equation}
and the Dirac velocity operator has the form
\begin{equation}
\bm v_D\equiv\frac{d\bm r}{dt}=i[{\cal H}_D,\bm r]=\bm\alpha.
\label{Diracvlct}
\end{equation}

The operator $\bm v_D$ is time-dependent:
\begin{equation}
\frac{d\bm v_D}{dt}=i[{\cal H}_D,\bm v_D]=i\{\bm\alpha,{\cal H}_D\}-2i\bm\alpha{\cal H}_D=2i(\bm p-\bm\alpha{\cal H}_D).
\label{Diraccel}
\end{equation}

The problem is usually considered in the Heisenberg picture:
\begin{equation}
\bm v_D(t)=e^{i{\cal H}_Dt}\bm\alpha e^{-i{\cal H}_Dt}.
\label{Heispte}
\end{equation}
In the Schr\"{o}dinger picture, the result is the same.
We suppose that 
the eigenvalues of the momentum and Hamiltonian operators are $\bm{\mathfrak{p}}$ and $H$, respectively. In this case, Eq. (\ref{Diraccel}) can be presented in terms of the Dirac velocity operator:
\begin{equation}
\frac{d\bm v_D}{dt}=2i(\bm{\mathfrak{p}}-\bm v_D H).
\label{Dprvelo}
\end{equation}
Its integration shows that the Dirac velocity oscillates:
\begin{equation}
\bm v_D(t)=\left[\bm v_D(0)-\frac{\bm{\mathfrak{p}}}{H}\right]e^{-2iHt}+\frac{\bm{\mathfrak{p}}}{H}.
\label{Dirvele}
\end{equation}
The evolution of the Dirac position operator obtained from this equation is given by
\begin{equation}
\bm r_D(t)=\bm r_D(0)+\frac{\bm{\mathfrak{p}}t}{H}+\frac{i}{2H}\left[\bm v_D(0)-\frac{\bm{\mathfrak{p}}}{H}\right]\left(e^{-2iHt}-1\right).
\label{Dirpoev}
\end{equation}

A similar result has been obtained for a free scalar (spin-0) particle (see Ref. \cite{ZitterbewegungFF} and references therein).
In this case, the initial Feshbach-Villars (FV) Hamiltonian reads \cite{FV}
\begin{equation}
{\cal H}_{FV}=\rho_3m+\left(\rho_3+i\rho_2\right)\frac{\bm p^2}{2m},
\label{HamFV}
\end{equation} where $\rho_i~(i=1,2,3)$ are the Pauli matrices.
The velocity operator in the FV representation is equal to
\begin{equation}
\bm v_{FV}=\left(\rho_3+i\rho_2\right)\frac{\bm p}{m}.
\label{velocFV}
\end{equation}
The corresponding acceleration operator is defined by the equation similar to Eq.
(\ref{Diraccel}) \cite{ZitterbewegungFF}:
\begin{equation}
\frac{d\bm v_{FV}}{dt}=i[{\cal H}_{FV},\bm v_{FV}]=i\{\bm v_{FV},{\cal H}_{FV}\}-2i\bm v_{FV}{\cal H}_{FV}=2i(\bm p-\bm v_{FV}{\cal H}_{FV}).
\label{FVaccel}
\end{equation} It is supposed that 
the eigenvalues of the momentum and Hamiltonian operators are $\bm{\mathfrak{p}}$ and $H$, respectively.
As a result, the final equations of dynamics of the free scalar particle \cite{ZitterbewegungFF} are equivalent to the corresponding equations for the
Dirac particle:
\begin{equation}
\bm v_{FV}(t)=\left[\bm v_{FV}(0)-\frac{\bm{\mathfrak{p}}}{H}\right]e^{-2iHt}+\frac{\bm{\mathfrak{p}}}{H},
\label{FVe}
\end{equation}
\begin{equation}
\bm r_{FV}(t)=\bm r_{FV}(0)+\frac{\bm{\mathfrak{p}}t}{H}+\frac{i}{2H}\left[\bm v_{FV}(0)-\frac{\bm{\mathfrak{p}}}{H}\right]\left(e^{-2iHt}-1\right).
\label{FVpoe}
\end{equation}

Zitterbewegung can also be considered in the GFV representation allowing one to describe not only massive scalar particles but also massless ones. This representation which has
been previously used in Refs. \cite{TMP2008,Honnefscalar,Honnefscalarnew,HonnefscalarLT} is, in fact, an infinite
set of representations. It has been obtained in Ref. \cite{ZitterbewegungPPNL} that equations of motion in the GFV representation 
are equivalent to Eqs. (\ref{FVe}), (\ref{FVpoe}). 

Amazingly, the same physical situation takes place for Proca (spin-1) particles. Massive Proca particles have been investigated in Refs. \cite{ZitterbewegungPPNL,ZitterbewegungJPhysConf}. Zitterbewegung takes place in the Sakata-Taketani representation
\cite{SaTa,SpinunitEDM} which does not exist for massless particles.
The final equations of motion of a free Proca particle are equivalent to the corresponding
equations for the Dirac and scalar particles \cite{ZitterbewegungPPNL,ZitterbewegungJPhysConf}:
\begin{equation}
\bm v_{ST}(t)=\left[\bm v_{ST}(0)-\frac{\bm{\mathfrak{p}}}{H}\right]e^{-2iHt}+\frac{\bm{\mathfrak{p}}}{H},
\label{STe}
\end{equation}
\begin{equation}
\bm r_{ST}(t)=\bm r_{ST}(0)+\frac{\bm{\mathfrak{p}}t}{H}+\frac{i}{2H}\left[\bm v_{ST}(0)-\frac{\bm{\mathfrak{p}}}{H}\right]\left(e^{-2iHt}-1\right).
\label{STpoe}
\end{equation}
In connection with this equivalence, we can mention
the existence of bosonic symmetries of the standard Dirac equation \cite{Simulik,Simulik1,Simulik2,Simulik3,Simulik4,Simulik5,Simuliknew}.

However, a coincidence of results obtained for particles with different spins in various representations does not mean that the effect of Zitterbewegung is observable. It has been pointed out in Ref. \cite{OConnell} that the transition to the FW representation establishes the proportionality of the operators $\bm p$ and $\bm v$ which should take place for free particles with any spin. In the FW representation, the acceleration vanishes and the Dirac Hamiltonian takes the form \cite{FW}
\begin{equation}
{\cal H}_{FW}=\beta\sqrt{m^2+\bm p^2}, \qquad \bm p\equiv-i\hbar\frac{\partial}{\partial\bm r}.
\label{HamFWDp}
\end{equation}
The velocity operator is given by
\begin{equation}
\bm v_{FW}=\beta\frac{\bm p}{\sqrt{m^2+\bm p^2}}=\frac{\bm p}{{\cal H}_{FW}}.
\label{eqvelmmD}
\end{equation} As a result, $d\bm v_{FW}/(dt)=0$ and Zitterbewegung does not take place. Similar relations have been obtained for massive particles with the spins 0 and 1 (see Ref. \cite{ZitterbewegungPPNL} and references therein).

The FW representation is the only one in which relativistic QM takes a Schr\"{o}dinger form and expectation values of all operators correspond to respective classical variables (see Refs. \cite{PRAFW,JMP,JINRLett12,ExpectationValue,relativisticFW} and references therein). Therefore, this representation is very convenient for checking an observability of physical effects. Zitterbewegung takes place \emph{only} for operators which are not the quantum-mechanical counterparts of the classical position and velocity. In particular, its appearance in the Dirac representation is caused by a significant difference between physical meanings of the Dirac and FW position operators (see Ref. \cite{PRAFW} and references therein).

\section{Distinguishing features of a quantum-mechanical description of massless particles} \label{masslessQM}

It should also be taken into account that particles can be in localized and delocalized states (see Ref. \cite{Anderson}). Photons being quanta of electromagnetic waves are usually delocalized and are commonly described in the framework of the wave theory. This theory determines \emph{local} velocities of the wave field. 
Certainly, phase and group velocities are different.  A local phase velocity is
defined by the phase front $\Phi(\bm r)$, $v_p=\omega/|\nabla\Phi(\bm r)|$,
where $\omega=ck$ is the angular frequency \cite{Born-Wolf,Bouchard}.
Another frequently used formula for the local phase velocity has been obtained in Ref. \cite{Chen}
(see also Ref. \cite{HuangWuHu}):
\begin{equation}
v_p=c\left[1+\frac{\nabla^2{\cal A}(\bm r)}{k^2{\cal A}(\bm r)}\right]^{-1/2}.
\label{LPV}
\end{equation}
The local group velocity is given by $v_g=|\partial_\omega\nabla\Phi(\bm r)|^{-1}$
\cite{Born-Wolf,Bouchard} (see also Ref. \cite{Saari} for details).
The both local velocities can be subluminal and superluminal depending on a region.
Certainly, they characterize important properties of light beams. For example, the local phase velocity defines an electron acceleration in a
laser beam \cite{WangScheidHo,ZhangWangXie,WangHoYuan}. The distribution of the local phase velocity
has been measured in Ref. \cite{WangQian}.

The quantum-mechanical approach substantially differs. It is known that the photon and other massless particles cannot be spatially localized \cite{LandauPeierls,NewtonWigner,Wightman,Ali}. In any case, this statement is valid for plane-wave states. Conditions admitting a localization of massless particles have been considered in Ref. \cite{Ingall}. Examples of light fields localized in two dimensions are Hermite-Gaussian, Laguerre-Gaussian, and Gaussian beams \cite{footnote}. Amazingly, the two-dimensional spatial localization of structured light beams results in nonzero \emph{effective} masses of their quanta \cite{LightArXiv}.

As a result, the position operator
$\bm r=(x,y,z)$ does not characterize the coordinates of massless particles. To determine a physical sense of this operator, 
we can use the classical Hamilton equation for the velocity:
\begin{equation}
\bm v=\frac{d\bm r}{dt}=\frac{\partial{\cal H}}{\partial\bm p},
\label{eqnHmln}
\end{equation} where ${\cal H}$ is the Hamiltonian. In QM, this equation remains valid as an operator equation and $\partial{\cal H}/(\partial\bm p)=i[{\cal H},\bm r]$. For massless particles, ${\cal H}$ and $\bm p$ correspond to $\hbar\omega$ and $\hbar\bm k$, respectively, and $\partial{\cal H}/(\partial\bm p)$ is the group velocity. The phase velocity is equal to $v_{ph} =\omega/k$. Thus, $\bm r$ is a radius vector of \emph{a moving point characterizing any fixed point of the envelope of the wave packet} and $\bm v$ is the velocity of its motion. The latter quantity defines the group velocity of a massless particle \emph{as a whole}. Despite great achievements of the wave theory, it cannot rigorously describe the motion of field quanta because any field quantum is extended over the entire space. In particular, a rigorous analysis of a subluminality of Gaussian and other structured light beams has been recently carried out just in the framework of QM \cite{LightArXiv}.

The descriptions of massless and massive particles in QM are substantially different. In the general case, basic equations for massless particles cannot be obtained from the corresponding equations for massive ones either by the substitution $m=0$ or in the limit of $m\rightarrow0$.  
    First of all, the definition of the spin should be radically changed. For massive particles, the conventional three-component spin (pseudo)vector $\bm s$ is defined in the particle rest frame (see Ref. \cite{PRAFW} and references therein). However, such a frame does not exist for massless particles. For such particles, the spin can be introduced but its projection onto the momentum direction can have only two values, minimum and maximum. Thus, the helicity of a massless particle is equal to (see, e.g., Ref. \cite{BLP})
\begin{equation}
h=\frac{\bm s\cdot\bm p}{p}=\pm s,
\label{helicit}
\end{equation} where $s$ is the spin quantum number.
Two partial waves describing states with $h=\pm s$ can be coherent when these states have the same energy. 
When the helicity is fixed, this polarization is circular. When the two partial waves describing states with the opposite helicity are coherent, the particle polarization is different (e.g., it can be linear).

It can be easily shown that the Dirac-Pauli spin algebra is applicable for a description of massless Dirac fermions. Squaring Eq. (\ref{helicit}) 
results in 
\begin{equation}
(\bm s\cdot\bm p)^2=\frac14\bm p^2.
\label{eleqvelmmD}
\end{equation} This equation is satisfied with the Pauli matrix $\bm \sigma=2\bm s$. Thus, the polarization of massless Dirac fermions can be defined by the conventional Pauli and Dirac matrices. As a result, the Dirac equation with $m=0$ can be used for a description of massless fermions. Of course, the operators $\bm\sigma$ and $\bm\Sigma$ characterize the polarization and helicity of massless particles but do not define the spin in the particle rest frame.

The situation is 
different for particles with higher spins including the photon ($s=1$) and the graviton ($s=2$). 
The number of components of wave
functions is defined by the number of independent spin components ($2s+1$ for massive particles). The quantum-mechanical description is equally applicable to states with a positive and a negative total energy. For massive particles, the minimum number of components of wave
functions is therefore equal to $2(2s+1)$. Such wave
functions are bispinors for Dirac particles and bispinor-like wave
functions for particles with other spins. In particular, the bispinor-like wave
functions have been successfully used for massive spin-1 particles \cite{ZitterbewegungPPNL,SpinunitEDM,YB,spin1}.

Some \emph{initial} equations describing spinning massive particles may be applicable to massless ones. However, the condition (\ref{helicit}) defining the helicity should be satisfied in the latter case. This condition defines two admissible longitudinal spin projections and reduces the number of independent spinor-like wave functions to four. In addition, wave functions of massless particles loose the probabilistic interpretation and should be considered as a distribution of a particle field strength. We can conclude that Hamiltonian equations for massive and massless particles substantially differ and massless particles should be considered separately. This conclusion is valid even for $s=0,\,1/2$ while the Klein-Gordon and Dirac equations also cover massless particles. For massless scalar particles, the well-known FV transformation \cite{FV} becomes inapplicable and the GFV one \cite{TMP2008} should be used. For massless Dirac particles, one should take into account the condition (\ref{helicit}). However, one of properties of the Dirac-Pauli spin algebra is a possibility to present any spin polarization as a coherent superposition of two basic states with the helicity $h=\pm1/2$. Therefore, the case of massless Dirac fermions is not very special and can basically be obtained from previous results for massive spin-1/2 particles. Of course, a difference between photons and massive spin-1 particles is much more substantial.

Zitterbewegung is a rather important quantum-mechanical problem and massless particles occupy a significant place in QM. However, the problem of Zitterbewegung of massless particles was not appropriately studied in previous investigations. In particular, the existence of Zitterbewegung of free photons has been claimed \cite{Unal,Kobe}. It has been noted in Refs. \cite{ZhiYongWang1,ZhiYongWang2,ZhiYongWang3} that Zitterbewegung does not exist for free photons but takes place for interacting ones and is observable in this case. We consider Zitterbewegung of photons in Secs. \ref{photons}, \ref{External}.

Therefore, the problem of Zitterbewegung of massless particles is still unsolved. Its detailed study fulfilled in the present paper is rather important. 
In particular, the unsolved issue of great interest is Zitterbewegung in external fields.

\section{Zitterbewegung of massless scalar and Dirac particles}\label{ScalarD}

In this section, we show an absence of Zitterbewegung of massless scalar and Dirac particles in the FW representation and its existence in some other representations. 
We apply the conventional commutative spatial coordinates. The use of noncommutative coordinates (i.e., noncommutative geometry) has been considered, e.g., in Ref. \cite{PRAFW}.

Certainly, we can utilize the general equation for the Hamiltonian in the GFV representation derived in Ref. \cite{TMP2008}. For a massless scalar particle, the initial Klein-Gordon equation has the form
\begin{equation} \left(\frac{\partial^2}{\partial t^2}-\nabla^2\right)\psi=0
\label{eqKlGol}
\end{equation}
and the GFV Hamiltonian is given by
\begin{equation}
{\cal H}_{GFV}=\rho_3\frac{\bm p^2+N^2}{2N}+i\rho_2\frac{\bm p^2-N^2}{2N},
\label{HamnGFV}
\end{equation} where $N$ is an arbitrary real nonzero parameter. Hereinafter,  $\rho_1,\rho_2$, and $\rho_3$ are the Pauli matrices: 
\begin{equation} \rho_1=\left(\begin{array}{cc} 0 & 1 \\ 1 & 0 \end{array}\right),\qquad
\rho_2=\left(\begin{array}{cc} 0 & -i \\ i & 0 \end{array}\right),\qquad \rho_3=\left(\begin{array}{cc} 1 & 0 \\ 0 & -1 \end{array}\right).  \label{Paulima} \end{equation}
For a scalar particle, the normalization of the two-component wave function in the FV representation \cite{FV}
\begin{equation} \Psi=\left(\begin{array}{c} \phi \\ \chi \end{array}\right) \label{wavefnct} \end{equation}
is given by $$\int{\Psi^\dag\rho_3\Psi dV}=1.
$$
Any GFV Hamiltonian is pseudo-Hermitian
(more exactly, $\rho_3$-pseudo-Hermitian): ${\cal H}_{GFV}^\ddag=\rho_3{\cal H}_{GFV}^\dag\rho_3={\cal H}_{GFV}$. 
For massless and massive particles, the normalization of GFV and FV wave functions is the same:
$$\int{\Psi_{GFV}^\dag\rho_3\Psi_{GFV} dV}=1.
$$ 

The velocity operator in the GFV representation is equal to
\begin{equation}
\bm v_{GFV}=\left(\rho_3+i\rho_2\right)\frac{\bm p}{N}.
\label{veloGFV}
\end{equation}
The corresponding acceleration operator reads
\begin{equation}
\frac{d\bm v_{GFV}}{dt}=i[{\cal H}_{GFV},\bm v_{GFV}]=i\{\bm v_{GFV},{\cal H}_{GFV}\}-2i\bm v_{GFV}{\cal H}_{GFV}=2i(\bm p-\bm v_{GFV}{\cal H}_{GFV}).
\label{GFVacce}
\end{equation} 
The dynamics of the massless scalar particle is independent of $N$ and is defined by the following equations:
\begin{equation}
\bm v_{GFV}(t)=\left[\bm v_{GFV}(0)-\frac{\bm{\mathfrak{p}}}{H}\right]e^{-2iHt}+\frac{\bm{\mathfrak{p}}}{H},
\label{GFVe}
\end{equation}
\begin{equation}
\bm r_{GFV}(t)=\bm r_{GFV}(0)+\frac{\bm{\mathfrak{p}}t}{H}+\frac{i}{2H}\left[\bm v_{GFV}(0)-\frac{\bm{\mathfrak{p}}}{H}\right]\left(e^{-2iHt}-1\right).
\label{GFVpoeq}
\end{equation} These equations are the same as the equations previously derived for massive particles.

For a massless Dirac particle, we use the initial Dirac equation (\ref{HamD}), suppose that $m=0$, and repeat all calculations presented in Sec. \ref{Previous}. Dynamic equations are the same as Eqs. (\ref{Dirvele}) and (\ref{Dirpoev}) and show the existence of Zitterbewegung.

As well as for massive particles, Zitterbewegung does not appear in the FW representation. The use of this representation for spinning massless 
particles should be commented. 



For a massless Dirac particle, the FW transformation operator is given by (cf. Refs. \cite{FW,JMP,PRA2015})
\begin{equation}
U_{FW}=\frac{p+\bm\gamma\cdot\bm p}{\sqrt{2}p}.
\label{eqFWfree}
\end{equation}
The FW Hamiltonian reads
\begin{equation}
{\cal H}_{FW}=\beta\sqrt{\bm p^2}=\beta p
\label{HamFWDn}
\end{equation}
and the velocity operator is given by
\begin{equation}
\bm v_{FW}=\beta\frac{\bm p}{p}=\frac{\bm p}{{\cal H}_{FW}}.
\label{eqvelmlD}
\end{equation}

Thus, the velocity and momentum are proportional. We can repeat the conclusion \cite{Zitterbewegungbook,ZitterbewegungOC}
that Zitterbewegung is the result of the interference between positive and negative energy states. In the FW representation, it disappears not only for massive \cite{ZitterbewegungKrekora,Zitterbewegungbook,ZitterbewegungOC} but also for massless Dirac particles.


\section{Quantum mechanics of the photon}\label{QMphotn}

While the quantum theory of radiation is well known \cite{Hejtler} and QM of the photon has a long history (see Refs. \cite{DarwinC,MohrP,BialBirOpt} and references therein), some important results, in particular, the FW transformation for the photon \cite{Barnett-FWQM}, have been obtained comparatively recently. Photon (light) beams are also extensively studied in optics. In optics, the wave function of the photon, $\Psi$, is not a wave function in the same sense as for the electron and determines the relative amplitude of the electric field \cite{Allen,Siegman,BialBirOpt}. 
The full description of an electromagnetic field including its interaction with matter is based on the quantum field theory (see Refs. \cite{Loudon,Srednicki}). However, the propagation of light in a free space can be adequately described with the Riemann-Silberstein vector $$\bm F=\frac{1}{\sqrt2}\left(\bm E+ i\bm B\right).$$ It 
allows one to reduce the Maxwell equations and to present them in the form
\cite{BialynickiBirulaPhysScr,Dirac-like}
\begin{equation}
i\hbar\frac{\partial\bm F}{\partial t}=c(\bm S\cdot\bm p)\bm F, \label{Weyllike}
\end{equation} where $\bm S=(S_1,S_2,S_3)$ is a vector in which the components are the conventional spin-1 matrices
\cite{YB}:
\begin{equation}
S_{1}=\left(\begin{array}{ccc} 0& 0& 0 \\ 0& 0 & -i \\ 0 & i & 0 \end{array}\right), \quad
S_{2}=\left(\begin{array}{ccc} 0& 0& i \\ 0& 0 & 0 \\ -i & 0 & 0 \end{array}\right), \quad
S_{3}=\left(\begin{array}{ccc} 0& -i & 0 \\ i & 0 & 0 \\ 0 & 0 & 0 \end{array}\right).
\label{spinunitmatr}
\end{equation}
This definition is not unique. One can use any other spin matrices satisfying the properties
\begin{equation}
[S_{i},S_{j}]=ie_{ijk}S_{k}, \quad
S_{i}S_{j}S_{k}+S_{k}S_{j}S_{i}=\delta_{ij}S_{k}+\delta_{jk}S_{i},\quad\bm S^2=2{\cal
I},
\label{spinmatrprop}
\end{equation} where ${\cal I}$ is the unit $3\times3$ matrix. The spin matrices act on three components of $\bm F$. The equation (\ref{Weyllike}) is similar to the Weyl equation for a massless Dirac particle \cite{Dirac-like}.
When the six-component wave function is defined by \cite{Barnett-FWQM,DarwinC,MohrP}
\begin{equation}
\Psi=\frac{1}{\sqrt2}\left(\begin{array}{c} \bm\phi \\ \bm\chi \end{array}\right)\equiv\frac{1}{\sqrt2}\left(\begin{array}{c} \bm E \\ i\bm B \end{array}\right), \label{wavefnt}
\end{equation}
the Dirac-like equation for the free electromagnetic field can be obtained \cite{Barnett-FWQM,DarwinC,MohrP}:
\begin{equation}
i\hbar\frac{\partial\Psi}{\partial t}=\bm\alpha\cdot\bm p\,\Psi, \qquad\bm\alpha=\left(\begin{array}{cc} 0 & \bm S \\ \bm S & 0 \end{array}\right).\label{Diraclike}\end{equation} In this equation, ${\cal H}=\bm\alpha\cdot\bm p$ is the Dirac-like Hamiltonian. Only Eq. (\ref{Diraclike}) has been taken into account in Ref. \cite{ZhiYongWang2}.

The additional condition of an orthogonality of the momentum direction to the fields $\bm E$ and $\bm B$ should also be taken into account. It follows from the Maxwell equations in the free space that $\bm p\cdot\bm\phi=\bm p\cdot\bm\chi=0$. As a result, the number of independent components of $\Psi$ reduces to four.

The FW transformation of Eq. (\ref{Diraclike}) has been carried out only in 2014 by Barnett \cite{Barnett-FWQM}. It is instructive to mention that any operator $\bm V$ satisfies the relation
$$i\bm V\times\bm G=(\bm S\cdot\bm V)\bm G,$$
where $\bm G$ is equal to $\bm E$ or $\bm B$. The FW transformation operator is equivalent to the operator (\ref{eqFWfree}) for the massless Dirac particle but has six components:
\begin{equation}
U_{FW}=\frac{p+\beta\bm\alpha\cdot\bm p}{\sqrt{2}p},\qquad \beta=\left(\begin{array}{cc} {\cal I}& 0 \\ 0& -{\cal I} \end{array}\right).
\label{eqFWpho}
\end{equation}

As a result, the FW wave function of the photon, $\Psi_{FW}=U_{FW}\Psi$, is defined by 
\begin{equation}\begin{array}{c} 
\Psi_{FW}^{(+)}=\left(\begin{array}{c} \bm \phi_{FW}^{(+)} \\ 0 \end{array}\right)=\left(\begin{array}{c} \bm E \\ 0 \end{array}\right) \quad {\rm if}\quad H=|\bm{\mathfrak{p}}|>0,\\ 
\Psi_{FW}^{(-)}=\left(\begin{array}{c} 0 \\ i\bm \chi_{FW}^{(-)} \end{array}\right)=\left(\begin{array}{c} 0 \\ i\bm B \end{array}\right) \quad {\rm if}\quad H=-|\bm{\mathfrak{p}}|<0.
\end{array}\label{eqFWwfn}
\end{equation} Here $\bm{\mathfrak{p}}$ and $H$ are eigenvalues of the momentum and Hamiltonian operators, respectively. Negative values of $H$ define photon states with a negative total energy. In Eq. (\ref{eqFWwfn}), the fields $\bm E$ and $\bm B$ also characterize states with a positive and a negative total energy, respectively. It is convenient to present the fields in the matrix form, $\bm E=\left(\begin{array}{c} E_1\\ E_2 \\ E_3\end{array}\right)$ and $\bm B=\left(\begin{array}{c} B_1\\ B_2 \\ B_3\end{array}\right)$. 

For massless particles including the photon, the physical meaning of negative-energy states is mostly the same as for massive particles. The problem of these states cannot be satisfactorily resolved in the context of relativistic QM \cite{Sachs} but the negative-energy states are rather important for quantum electrodynamics and quantum field theory \cite{Feynman}. Such states can be used for a description of virtual particles and a state with a negative energy can characterize a virtual photon. We should note that masses of particles with negative energies are also negative \cite{Debergh}. Some of the most important forces can be considered as an exchange of virtual particles. This possibility exists for all major forces like electromagnetic, weak, strong, and gravitational ones. In particular, the electromagnetic interaction of a charged body can be interpreted as an exchange of virtual photons. Virtual photons play a pivotal role in quantum physics (see, e.g., Refs. \cite{ZhiYongWang1,ZhiYongWang2,ZhiYongWang3,Wolf,Mart,Liu,Liberato,Alekseev}).

In the FW representation, $\bm p\cdot\bm\phi=0$ if $H=|\bm{\mathfrak{p}}|>0$ and $\bm p\cdot\bm\chi=0$ if $H=-
|\bm{\mathfrak{p}}|<0$. Therefore, the number of nonzero and independent components of $\Psi_{FW}$ reduces to two.
It has been proven in Ref. \cite{Barnett-FWQM} that 
\begin{equation} {\cal P}^2\Psi\equiv[\bm p^2-(\bm\alpha\cdot\bm p)^2]\Psi=[\bm p^2-(\bm\Sigma\cdot\bm p)^2]\Psi=0, \qquad \bm\Sigma=\left(\begin{array}{cc} \bm S & 0 \\ 0& \bm S \end{array}\right).
\label{eqnFWcp}\end{equation} 

As a result of derivations \cite{Barnett-FWQM}, the FW Hamiltonian is defined by
\begin{equation} i\hbar\frac{\partial\Psi_{FW}}{\partial t}={\cal H}_{FW}\Psi_{FW},\qquad {\cal H}_{FW}=\beta p.
\label{eqFWHfn}
\end{equation}

Quantum mechanics of the photon agrees with the general condition (\ref{helicit}) for the helicity. For the photon, $\bm s=\bm S$ and $\bm s=\bm\Sigma$ when one uses two-component and four-component wave functions, respectively, and $s=1$. Squaring Eq. (\ref{helicit}) leads to the relation $(\bm\Sigma\cdot\bm p)^2=p^2$ which is equivalent to Eq. (\ref{eqnFWcp}) (see Ref. \cite{Barnett-FWQM} and references therein). 

Equation (\ref{eqFWwfn}) does not mean that the electric and magnetic fields defines the FW wave functions of real and virtual photons, respectively. In the electromagnetic wave, these fields are not independent and are connected by the relations $\bm B=\bm n\times\bm E,~\bm E=-\bm n\times\bm B,~\bm n=\bm{\mathfrak{p}}/|\bm{\mathfrak{p}}|$. In addition, one can introduce the transformed spinor-like wave functions ${\bm \phi'}_{FW}^{(+)}=\bm n\times{\bm \phi}_{FW}^{(+)},~{\bm \chi'}_{FW}^{(-)}=\bm n\times{\bm \chi}_{FW}^{(-)}$ with the same FW Hamiltonian (\ref{eqFWHfn}). As a result, the transformed FW wave functions take the form
\begin{equation}\begin{array}{c} 
{\Psi'}_{FW}^{(+)}=\left(\begin{array}{c} {\bm \phi'}_{FW}^{(+)} \\ 0 \end{array}\right)=\left(\begin{array}{c} \bm B \\ 0 \end{array}\right) \quad {\rm if}\quad H=|\bm{\mathfrak{p}}|>0,\\ 
{\Psi'}_{FW}^{(-)}=\left(\begin{array}{c} 0 \\ i{\bm \chi'}_{FW}^{(-)} \end{array}\right)=\left(\begin{array}{c} 0 \\ -i\bm E \end{array}\right) \quad {\rm if}\quad H=-|\bm{\mathfrak{p}}|<0.
\end{array}\label{eqFWtfn}
\end{equation}
Equations (\ref{eqnFWcp}) and (\ref{eqFWtfn}) show that the both fields can be equivalently used for a definition of real and virtual photons.

To complete QM of the photon, we need to determine an equation of the second order in the temporal and spatial derivatives (Klein-Gordon-like equation). The quantum-mechanical description of the photon in the GFV representation will be carried out in Sec. \ref{arbitrary}. The Klein-Gordon-like (KGL) equation can be easily obtained by squaring the FW Hamiltonian equation (\ref{eqFWHfn}) and has the form
\begin{equation} \left(\frac{\partial^2}{\partial t^2}-\nabla^2\right)\bm\psi=0.
\label{eqKlGor}
\end{equation}
The three-component wave function $\bm\psi$ coincides with the upper three-component spinor-like part of $\Psi_{FW}$ and is equal to $\bm E$ for positive-energy states of the photon. For negative-energy states, $\bm\psi=i\bm B$. The number of independent components of $\psi$ reduces to two due to the orthogonality condition $\bm p\cdot\bm\psi=0$.

It can also be shown by squaring Eq. (\ref{Weyllike}) and taking into account Eqs. (\ref{helicit}) and (\ref{eqnFWcp}) that the Riemann-Silberstein vector is also a solution of the KGL equation, $\bm\psi=\bm F=\frac{1}{\sqrt2}\left(\bm E+ i\bm B\right)$, for states with both positive and negative total energies.

We note the nonequivalence of the Dirac-like and FW wave functions, $\Psi$ and $\Psi_{FW}$. The former function defines both fields, $\bm E$ and $\bm B$, reproduces the Maxwell equations, and establishes the connection between states with a positive and a negative total energy. While the FW wave function perfectly describes the light field, it does not possesses these properties. However, just the wave function defined only by the electric field strength is used in optics.

\section{Zitterbewegung of the free photon}\label{photons}

The use of the Dirac-like equation (\ref{Diraclike}) and the KGL equation (\ref{eqKlGor}) allows us to study Zitterbewegung of the photon. The velocity and acceleration operators determined from Eq. (\ref{Diraclike}) are given by
\begin{equation}
\bm v\equiv\frac{d\bm r}{dt}=i[{\cal H},\bm r]=\bm\alpha,\qquad
\frac{d\bm v}{dt}=i[{\cal H},\bm v]=i\{\bm\alpha,{\cal H}\}-2i\bm\alpha{\cal H}=2i(\bm p-\bm\alpha{\cal H})
\label{Diraclk}
\end{equation} and are equivalent to the corresponding equations (\ref{Diracvlct}) and (\ref{Diraccel}) for Dirac particles. Certainly, the derivation for the photon leads to the final formulas which are also equivalent to the corresponding ones for Dirac particles:
\begin{equation} 
\bm r(t)=\bm r(0)+\frac{\bm{\mathfrak{p}}t}{H}+\frac{i}{2H}\left[\bm v(0)-\frac{\bm{\mathfrak{p}}}{H}\right]\left(e^{-2iHt}-1\right),\label{DirlZbe}
\end{equation}
\begin{equation}
\bm v(t)=\left[\bm v(0)-\frac{\bm{\mathfrak{p}}}{H}\right]e^{-2iHt}+\frac{\bm{\mathfrak{p}}}{H}.
\label{DirlZbv}
\end{equation}

To study Zitterbewegung, we can also use the GFV transformation of the KGL equation (\ref{eqKlGor}). In this case, we introduce the spinor-like wave functions (cf. Ref. \cite{TMP2008})
\begin{equation} \bm\psi=\bm\phi+\bm\chi, \qquad
i\frac{\partial\bm\psi}{\partial t}= N(\bm\phi-\bm\chi).
\label{eqFVT}\end{equation}
Multiplying the second equation by $i\partial/(\partial t)$ allows one to obtain the GFV Hamiltonian \cite{TMP2008}:
\begin{equation}
{\cal H}_{GFV}=\rho_3\frac{\bm p^2+N^2}{2N}+i\rho_2\frac{\bm p^2-N^2}{2N}.
\label{HamnGFVo}
\end{equation} We should mention that this Hamiltonian is also proportional to the $3\times3$ unit matrix which is omitted (${\cal H}_{GFV}{\cal I}\rightarrow{\cal H}_{GFV}$). This is its only difference with the Hamiltonian (\ref{HamnGFV}). As a result, the final equations describing Zitterbewegung coincide with Eqs. (\ref{GFVe}) and (\ref{GFVpoeq}) for the scalar particle and with Eqs. (\ref{DirlZbe}) and (\ref{DirlZbv}) in the Dirac-like representation.

Thus, Zitterbewegung of the photon in the Dirac-like and GFV representations is determined by the same equations which also coincide with all corresponding equations for other massless and massive particles presented in Secs. \ref{Previous}--\ref{ScalarD}. Nevertheless, all noted equations have been obtained in representations different from the FW representation. The transition to the latter representation resulting in the Hamiltonian (\ref{eqFWHfn}) eliminates Zitterbewegung. In this representation, the velocity operator has the form [cf. Eqs. (\ref{eqvelmmD}) and (\ref{eqvelmlD})]
\begin{equation}
\bm v_{FW}=\beta\frac{\bm p}{\sqrt{m^2+\bm p^2}}=\frac{\bm p}{{\cal H}_{FW}}.
\label{eqvelmmp}
\end{equation} Since the FW position and velocity operators are the quantum-mechanical counterparts of the corresponding classical variables (see Ref. \cite{PRAFW} and references therein), the observability of Zitterbewegung should be determined just in the FW representation. As a result of proportionality of the velocity and momentum operators, Zitterbewegung of the free photon is unobservable. Such Zitterbewegung should be considered as a purely mathematical effect and does not exist as a real physical one. These our conclusions contradict to the conclusions made in Refs. \cite{Unal,Kobe}. 

\section{Zitterbewegung of particles with an arbitrary spin in the generalized Feshbach-Villars representation}\label{arbitrary}

Zitterbewegung can also be studied for particles with an arbitrary spin. It has been shown by Weinberg \cite{Weinberg} that particles with an arbitrary spin satisfy the Klein-Gordon (more precisely, Klein-Gordon-like) equation
\begin{equation} \left(\frac{\partial^2}{\partial t^2}-\nabla^2+m^2\right)\psi=0,
\label{eqKlGorlik}
\end{equation}
where the wave function $\psi$ has $2s+1$ components. The number of components of this wave function is defined by the number of independent spin components. Evidently, Eq. (\ref{eqKlGorlik}) is equally applicable to states with a positive and a negative total energy. Any passage to the Hamiltonian formalism joins such states and needs an introduction of the $2(2s+1)$-component wave function $\Psi$.

Equation (\ref{eqKlGorlik}) is also applicable to massless particles. However, the condition (\ref{helicit}) defining the helicity should be satisfied in this case. This condition defines two admissible longitudinal spin projections and reduces the number of independent components of $\psi$ to two.

We can carry out the GFV transformation \cite{ZitterbewegungPPNL,TMP2008}
\begin{equation}
\psi=\phi+\chi,\qquad i\frac{\partial\psi}{\partial t} =N(\phi-\chi)
\label{KGGFV}
\end{equation}
for particles with an arbitrary spin. 

After multiplying the last relation by $i\partial/(\partial t)$, Eq. (\ref{KGGFV}) can be presented in the matrix form \cite{ZitterbewegungPPNL,TMP2008}
\begin{equation}\begin{array}{c}
 i\frac{\partial\Psi_{GFV}}{\partial t} ={\cal H}_{GFV}\Psi_{GFV},\qquad\Psi_{GFV}=\left(\begin{array}{c} \phi \\ \chi \end{array}\right),\\
{\cal H}_{GFV}=\rho_3\frac{\bm p^2+m^2+N^2}{2N}+i\rho_2\frac{\bm p^2+m^2-N^2}{2N}.
\end{array}\label{HamnGFm}
\end{equation} Here ${\cal H}_{GFV}$ and $\Psi_{GFV}$ are the Hamiltonian and the wave function in the GFV representation. Evidently, this representation connects the states with a positive and a negative total energy with each other. We note the difference between the ST and GFV Hamiltonians for massive spin-1 particles and underline the applicability of the GFV representation for both bosons and fermions. 
The same derivations as in Ref. \cite{ZitterbewegungPPNL} show that equations of motions in this representation coincide with the corresponding equations \cite{ZitterbewegungPPNL} for scalar particles and with Eqs. (\ref{GFVe}) and (\ref{GFVpoeq}). Thus, the description of particles with an arbitrary spin in the GFV representation demonstrates the presence of Zitterbewegung and the perfect similarity of equations of motion for particles with any spin in any representation (except for the FW one). The derivation is the same for massive and massless particles. In the latter case, all obtained formulas remain valid provided that $m=0$.

Like in other cases, Zitterbewegung does not take place in the FW representation. The FW transformation of the Hamiltonian (\ref{HamnGFm}) is exact. The general form of the FW transformation operator has been obtained in Ref. \cite{JMP}. In the considered case, this operator reduces to \cite{TMP2008}
\begin{equation}
U_{GFV\rightarrow FW}=\frac{\epsilon+N+\rho_1(\epsilon-N)}{2\sqrt{\epsilon N}},\qquad \epsilon=\sqrt{m^2+\bm p^2}.
\label{UGFVm}
\end{equation}
Since this operator is $\rho_3$-pseudounitary \cite{TMP2008}, the operator of the inverse transformation is defined by
\begin{equation} \begin{array}{c}
U_{GFV\rightarrow FW}^\dag=\rho_3U_{GFV\rightarrow FW}^{-1}\rho_3,\qquad U_{GFV\rightarrow FW}^{-1}=U_{FW\rightarrow GFV}=\rho_3U_{GFV\rightarrow FW}^\dag\rho_3,\\
U_{FW\rightarrow GFV}=\frac{\epsilon+N-\rho_1(\epsilon-N)}{2\sqrt{\epsilon N}}.
\end{array} \label{HamiGFV}
\end{equation}
For the photon, $m=0$ and the GFV wave function $\Psi_{GFV}=U_{FW\rightarrow GFV}\Psi_{FW}$ is equal to
\begin{equation} \begin{array}{c}
\Psi_{GFV}^{(+)}=\frac{\bm E}{2\sqrt{p N}}\left(\begin{array}{c} N+p \\ N-p \end{array}\right),\qquad
\Psi_{GFV}^{(-)}=\frac{i\bm B}{2\sqrt{p N}}\left(\begin{array}{c} N-p \\ N+p \end{array}\right).
\end{array} \label{wfGFV}
\end{equation} for states with a positive and a negative total energy, respectively. The FW wave function corresponding to these states is defined by Eq. (\ref{eqFWwfn}).

In the general case, we can determine the FW Hamiltonian and the corresponding wave function with the use of the operator (\ref{UGFVm}):
\begin{equation} \begin{array}{c}
{\cal H}_{FW}=\beta\epsilon,\qquad \Psi_{FW}=\Psi_{FW}^{(+)}=\frac{2\sqrt{\epsilon N}}{\epsilon+ N}\left(\begin{array}{c} \phi \\ 0 \end{array}\right) \quad {\rm if}\quad {\cal H}_{FW}=|\epsilon|>0,\\ \Psi_{FW}=
\Psi_{FW}^{(-)}=\frac{2\sqrt{\epsilon N}}{\epsilon+ N}\left(\begin{array}{c} 0 \\ \ \chi \end{array}\right) \quad {\rm if}\quad {\cal H}_{FW}=-|\epsilon|<0.
\end{array} \label{eqFWHwf}
\end{equation}

The connection between the FW wave function and the initial GFV one is similar to that between the FW and Dirac wave functions \cite{JINRLDFW} but the coupling factors are different in the two cases.

For massive and massless particles, the FW velocity operator is defined by 
\begin{equation}
\bm v_{FW}=i[{\cal H}_{FW},\bm r_{FW}]=\beta\frac{\bm p}{\sqrt{m^2+\bm p^2}}=\frac{\bm p}{{\cal H}_{FW}}.
\label{eqvellD}
\end{equation} This equation is valid for particles with any spin. For massless particles, $\bm v_{FW}=\beta c\bm p/|\bm p|$.

Therefore, $d\bm v_{FW}/(dt)=0$ and Zitterbewegung does not take place. Since the FW representation is the only representation in which relativistic QM takes a Schr\"{o}dinger form and expectation values of all operators correspond to respective classical variables, Zitterbewegung takes place \emph{only} for operators which are not the quantum-mechanical counterparts of the classical position and velocity. As a result, Zitterbewegung is not observable.  This is the same conclusion which has been made in the previous studies 
\cite{ZitterbewegungFF,ZitterbewegungPPNL,ZitterbewegungJPhysConf,ZitterbewegungKrekora,Zitterbewegungbook,ZitterbewegungOC,PRAFW,ZitterbewegungDeriglazov,ZitterbewegungKobakhidze}. 

\section{Zitterbewegung in external fields}\label{External}

The analysis of Zitterbewegung made in the above-mentioned previous publications is not excaustive because only free particles were studied. However, the problem of Zitterbewegung of particles in external 
fields is also very important. In particular, it has been stated in Refs. \cite{ZitterbewegungRo,ZitterbewegungBD,ZitterbewegungCapelle} that just Zitterbewegung causes the electrostatic contact (Darwin) interaction.

We will consider Zitterbewegung of particles with spins 0, 1/2, and 1 in external fields. Both massless and massive particles will be taken into consideration. This is necessary because a consideration of massive particles simplifies making a conclusion on an existence of Zitterbewegung in external fields. Furthermore, we cannot include the massless photon into our consideration because its electromagnetic interactions can be described by quantum field theory but not by QM.

We use the method of the relativistic 
FW transformation elaborated in Refs. \cite{JMP,ExpectationValue,PRA2015,PRA2008}. It is applicable for a particle in arbitrarily 
strong external fields (without taking into account effects of quantum field theory). We can mention that the Darwin interaction manifests itself even in the weak-field approximation. In this approximation, interactions with external fields do not 
significantly affect large Zitterbewegung in the Dirac and GFV representations and total Zitterbewegung is mostly defined by the equations for free particles (see Secs. 
\ref{Previous},\ref{ScalarD},\ref{photons},\ref{arbitrary}).
The problem of existence of Zitterbewegung in external fields is nontrivial because Zitterbewegung is absent for free particles in the FW representation (see Secs. \ref{Previous},\ref{ScalarD},\ref{photons},\ref{arbitrary}) but the Darwin interaction is usually explained by Zitterbewegung. 
To solve this problem, we 
use the previous derivations of relativistic FW Hamiltonians for scalar \cite{TMP2008} and spinning \cite{RPJ,PhysRevDspinunit} particles in electromagnetic fields. 

Certainly, the FW transformation is perturbative in the presence of
external fields and an exact FW
separation into positive- and negative-energy subspaces is not possible in a non-perturbative way. However,
the FW transformation method \cite{JMP,ExpectationValue,PRA2015,PRA2008} applied also in Ref. \cite{TMP2008} gives exact and compact relativistic expessions for all terms proportional to the
zero and first powers of the Planck constant and only for such terms proportional to $\hbar^2$ which describe contact interactions (including the Darwin one). Hereinafter, we will reproduce only such expressions and will omit other terms.
We should add that terms proportional to $\hbar$ describe spin interactions. While the precision of the used method of the FW transformation has been specified only in Ref. \cite{PRA2015}, previous results obtained by this method are also exact for the above-mentioned terms. The considered method of the FW transformation cannot be applied for a calculation of perturbations of next orders.

The relativistic FW Hamiltonian for a scalar particle in electromagnetic fields reads \cite{TMP2008}
\begin{equation}\begin{array}{c} {\cal H}_{FW}=\rho_3\epsilon'+ e\Phi, \qquad \epsilon'=\sqrt{m^2+\bm{\pi}^2}.
\end{array}\label{Peq15}\end{equation}
We underline the absence of terms describing contact interactions.

The corresponding FW Hamiltonian for a spin-1/2 particle with magnetic and electric dipole moments (MDM and EDM) is given by \cite{JMP,RPJ}
\begin{equation} {\cal H}_{FW}={\cal H}_{FW}^{(MDM)}+{\cal
H}_{FW}^{(EDM)}, \label{FWHamEDM} \end{equation}
\begin{equation}\begin{array}{c} {\cal H}_{FW}^{(MDM)}=\beta\epsilon'+e\Phi+\frac
   14\left\{\left(\frac{\mu_0m}{\epsilon'
   +m}+\mu'\right)\frac{1}{\epsilon'},\Bigl[\bm\Sigma\cdot(\bm\pi
\times\bm E-\bm E\times\bm\pi)-\hbar\nabla
\cdot\bm E\Bigr]\right\}\\ -\frac
12\left\{\left(\frac{\mu_0m}{\epsilon'}
+\mu'\right), \bm\Pi\!\cdot\!\bm B\right\}\\
+\beta\frac{\mu'}{4}\left\{\frac{1}{\epsilon'(\epsilon'+m)},
\Bigl[(\bm{B}\!\cdot\!\bm\pi)(\bm{\Sigma}\!\cdot\!\bm\pi)+ (\bm{\Sigma}
\!\cdot\!\bm\pi)(\bm\pi\!\cdot\!\bm{B})+2\pi\hbar(\bm\pi\!\cdot\!\bm j+
\bm j\!\cdot\! \bm\pi)\Bigr]\right\},
\end{array} \label{eq33new} \end{equation}
\begin{equation}
\begin{array}{c}
{\cal H}_{FW}^{(EDM)}=-d\bm\Pi\!\cdot\!\bm E
+\frac{d}{4}\left\{\frac{1}{\epsilon'(\epsilon'+m)},
\biggl[(\bm{E}\!\cdot\!\bm\pi)(\bm{\Pi}\!\cdot\!\bm\pi)+ (\bm{\Pi}
\!\cdot\!\bm\pi)(\bm\pi\!\cdot\!\bm{E})\biggr]\right\} \\-\frac
d4\left\{\frac{1}{\epsilon'},\biggl(\bm\Sigma\!\cdot\![\bm\pi\!
\times\!\bm B]-\bm\Sigma\!\cdot\![\bm
B\!\times\!\bm\pi]\biggr)\right\}.
\end{array} \label{EDMeq12} \end{equation} Here $\mu_0=e\hbar/(2mc)$ is the Dirac magnetic moment, $\mu'$ is the anomalous magnetic moment ($\mu=\mu_0+\mu'=ges\hbar/(2mc)$, where $g$ is the g-factor), $\epsilon'=\sqrt{m^2+\bm{\pi}^2}$, $\bm{\pi}=\bm p-e\bm A$, $\bm\Sigma$ and $\bm\Pi$ are Dirac matrices, $d$ is the electric dipole moment,
and
$$\bm j=\frac{1}{4\pi}\left(c\,\nabla\!\times\!\bm B-\frac{\partial \bm E}
{\partial t}\right)$$
is the density of external electric current. The term in Eq. (\ref{eq33new}) proportional to $\nabla
\cdot\bm E$ defines the Darwin (contact) interaction. While we take into account in Eq. (\ref{EDMeq12}) terms proportional to $\hbar^2$ and describing contact interactions with external charges and currents, such terms are zero due to the Maxwell equations
$$\nabla\cdot\bm B=0, \quad \nabla\times\bm E=-\frac{\partial \bm B}
{\partial t}.$$
Terms proportional to the second and higher
powers of $\hbar$ and quadratic/bilinear in $\bm E$ and $\bm B$ are neglected.

For a spin-1 particle with a magnetic moment, taking into account quadrupole interactions \cite{YB,JETP} leads to the following FW Hamiltonian \cite{JETP}:
\begin{equation} \begin{array}{c}
{\cal H}_{FW}=\rho_3\epsilon'+e\Phi+\frac{e}{4m}\Biggl[\left\{\left(\frac{g-2}{2}+
\frac{m}{\epsilon'+m}\right)\frac{1}{\epsilon'},
\left[\bm S\cdot(\bm\pi\times\bm E-\bm E\times
\bm\pi)\right]\right\}
 \\  -
\rho_3\left\{\left(g-2+\frac{2m}{\epsilon'}\right),
\bm S\cdot\bm B\right\}+\rho_3\left\{\frac{g-2}{2\epsilon'(\epsilon'+m)},
\{\bm S\cdot\bm\pi,\bm\pi\cdot\bm B\}\right\}\Biggr]
\\+\frac{e(g\!-\!1)}{4m^2}\biggl\{\Bigl(\bm S\cdot\nabla
-\frac{1}{\epsilon'(\epsilon'\!+\!m)}(\bm S\cdot\bm\pi)
(\bm\pi\cdot\nabla)\Bigr),\Bigl(\bm S\cdot\bm E-
\frac{1}{\epsilon'(\epsilon'\!+\!m)}(\bm S\cdot\bm\pi)
(\bm\pi\cdot\bm E)\Bigr)\biggr\}\\+
\frac{e}{8m^2}\left\{\frac{1}{\epsilon'(\epsilon'+m)}\left(g-1+\frac{m}{\epsilon'+m}\right),
\biggl\{\bm S\cdot[\bm\pi\times\nabla],\bm S\cdot[\bm\pi\times\bm E]\biggr\}\right\}\\-
\frac{e(g-1)}{2m^2}\nabla\cdot\bm
E+\frac{e}{4m^2}\left\{\frac{1}{{\epsilon'}^2}
\left(g-1+\frac{m^2}{4{\epsilon'}^2}\right),
(\bm\pi\cdot\nabla)(\bm\pi\cdot\bm E)\right\},
\end{array} \label{eq20nonun} \end{equation}
where the $g$ factor is used and the Planck constant is omitted. In this equation, a noncommutativity of operators is (partially) neglected. We should add that Eqs. (\ref{Peq15}) and (\ref{FWHamEDM})--(\ref{EDMeq12}) are appicable for massless particles while Eq. (\ref{eq20nonun}) describes only massive ones. 

The presented equations perfectly agree with corresponding classical equations describing the momentum and spin dynamics \cite{PRAFW,classEDM}. The agreement takes also place for a spin-1 particle with MDM and EDM \cite{PRAFW,SpinunitEDM}. 

Equations (\ref{Peq15})--(\ref{eq20nonun}) allow us to check the above-mentioned statement \cite{ZitterbewegungRo,ZitterbewegungBD,ZitterbewegungCapelle} that the electrostatic contact interaction (defined in the FW representation) is connected with Zitterbewegung. However, this statement cannot be right. 
The electrostatic contact interaction does not exist for scalar particles and exists for spinning ones while Zitterbewegung appears for any particles in representations different from the FW representation. An absence of observable Zitterbewegung for scalar particles and a similarity of Eqs. (\ref{Peq15})--(\ref{EDMeq12}) for massless and massive particles show that it not caused by the particle localization (established in Ref. \cite{NewtonWigner}) either. In fact, the electrostatic contact interaction is conditioned by the particle spin. Its essential difference for spin-1/2 and spin-1 partices confirms this important and unobvious statement. We can add that a spinning particle is not perfectly pointlike. In particular, its center of charge and center of mass do not coincide \cite{Pryce,SuttorpDeGroot,DeKerfBauerle,KhPom,PomKh,Obzor}.

The effective root-mean-square radii do not depend on external fields and are the same as in vacuum.

We repeat that an exact FW separation into positive- and negative-energy subspaces is not possible in a non-perturbative way and our conclusions cannot be extended to perturbations of next orders.

In this section, we have analyzed only Zitterbewegung which appears even for free particles (like a nonzero root-mean-square radius of the electron) but manifests itself only for interacting ones. None of leading terms in equations of motion describes such Zitterbewegung. However, it is evident that external 
fields can lead to a trembling motion (e.g., Zitterbewegung) of particles which is observable \emph{in these fields} but does not have any analogue for particles in free space. For example, a charge undergoes Zitterbewegung in a field of a coherent electromagnetic wave. It has been established in Refs. \cite{ZhiYongWang1,ZhiYongWang2,ZhiYongWang3} that Zitterbewegung takes place for \emph{real} photons interacting with \emph{virtual} longitudinal and scalar ones and this effect is observable. The effect cannot occur in the absence of virtual photons. It is described in the framework of quantum field theory and cannot be considered as a quantum-mechanical effect.

\section{Discussion and summary}\label{Summary}

We have analyzed Zitterbewegung of massless particles with an arbitrary spin in various representations. While Zitterbewegung of massless particles in the free space has been extensively studied, 
the quantum-mechanical descriptions of massless and massive particles substantially differ. In the general case, basic equations for massless particles cannot be obtained from the corresponding equations for massive ones and 
the definition of the spin should be substantially changed. For massive particles, the conventional three-component spin (pseudo)vector $\bm s$ is defined in the particle rest frame (see Ref. \cite{PRAFW} and references therein). However, such a frame does not exist for massless particles. For such particles, the helicity has only two admissible values, $+s$ and $-s$. The spin defines the polarization and helicity of massless particles and cannot be defined in the particle rest frame. If $s\neq0$, the bispinor-like wave functions have only four independent components. Therefore, Hamiltonian equations for massive and massless particles substantially differ and massless particles should be considered separately. For this reason, the results obtained are not obvious.

Photons and other field quanta are extended over the entire space and the radius vector $\bm r=(x,y,z)$ cannot define their coordinates. Therefore, $\bm r$ is a radius vector of a moving point characterizing any fixed point of the envelope of the wave packet characterizing the particle and $\bm v$ is the velocity of its motion. The latter quantity
defines the group velocity of the massless particle \emph{as a whole} but not that of any point of the wavefront. We have considered massless scalar and Dirac particles, the photon, and massive and massless particles with an arbitrary spin. We have described them in different representations. For particles with an arbitrary spin, the GFV and FW representations have been used.
In all cases, Zitterbewegung takes place in any representation except for the FW one. In such cases, formulas describing the particle motion are the same in any representation. In the FW representation, the operators of the velocity and momentum are proportional and Zitterbewegung does not take place. The radius vector (position) and velocity operators are the quantum-mechanical counterparts of the classical position and velocity just in the FW representation. Therefore, Zitterbewegung of free particles is not observable.  This conclusion agrees with the conclusions made in the previous studies 
\cite{ZitterbewegungFF,ZitterbewegungPPNL,ZitterbewegungJPhysConf,ZitterbewegungKrekora,Zitterbewegungbook,ZitterbewegungOC,PRAFW,ZitterbewegungDeriglazov,ZitterbewegungKobakhidze}
for massive particles. However, it disagrees with the corresponding conclusions reached in many other investigations including precedent papers devoted to Zitterbewegung of the free photon \cite{Unal,Kobe}.

The very important problem is Zitterbewegung of particles in external fields. Zitterbewegung has been noted in Refs. \cite{ZitterbewegungRo,ZitterbewegungBD,ZitterbewegungCapelle} as a reason of the electrostatic contact (Darwin) interaction being a manifestation of effective root-mean-square radii of spinning particles. However, our analysis given in Sec. \ref{External} shows that this interaction cannot be caused by Zitterbewegung and is conditioned by the particle spin. We underline that the effective root-mean-square radii do not depend on external fields and are the same as in vacuum. Since an exact FW transformation is not possible in a non-perturbative way, our conclusions are restricted by the applicability of the used transformation method and cannot be extended to perturbations of next orders.

We have analyzed only Zitterbewegung which appears even for free particles (like a nonzero root-mean-square radius of the electron) but manifests itself only for interacting ones. None of leading terms in equations of motion describes such Zitterbewegung. However, it is evident that external 
fields can lead to a trembling motion (e.g., Zitterbewegung) of particles which is observable \emph{in these fields} but does not have any analogue for particles in free space. For example, a charge undergoes Zitterbewegung in a field of a coherent electromagnetic wave. It has been established in Refs. \cite{ZhiYongWang1,ZhiYongWang2,ZhiYongWang3} that Zitterbewegung takes place for \emph{real} photons interacting with \emph{virtual} longitudinal and scalar ones and this effect is observable. The effect cannot occur in the absence of virtual photons. It is described in the framework of quantum field theory and cannot be considered as a quantum-mechanical effect.

Thus, Zitterbewegung of particles can appear due to interactions with external fields and effects of quantum field theory (see Refs. \cite{ZhiYongWang1,ZhiYongWang2,ZhiYongWang3}). In such cases, it can be observable.

\begin{acknowledgments}
This work was supported in part 
by the National Natural Science
Foundation of China (Grants No. 11575254 and No. 11805242), and
by the National Key Research and Development Program of China
(No. 2016YFE0130800).
%
The author also acknowledges hospitality and support by the
Institute of Modern
Physics of the Chinese Academy of Sciences. 
\end{acknowledgments}



\begin{thebibliography}{}

\bibitem{ZitterbewegungSc}
E. Schr\"{o}dinger, \"{U}ber die kr\"{a}ftefreie Bewegung in der relativistischen Quantenmechanik, Berlin Ber., 418 
(1930); Zur Quantendynamik des Elektrons, Berlin Ber., 63 
(1931).

\bibitem{ZitterbewegungGG}
R. F. Guertin and E. Guth, Zitterbewegung in Relativistic Spin-0 and -1/2 Hamiltonian Theories, Phys. Rev. D \textbf{7}, 1057 (1973).

\bibitem{ZitterbewegungFF}
M. G. Fuda and E. Furlani, Zitterbewegung and the Klein paradox for spinzero particles, Am. J. Phys. \textbf{50}, 545 (1982).

\bibitem{ZitterbewegungPPNL}
A. J. Silenko, Zitterbewegung of Bosons,  Phys. Part. Nucl. Lett. \textbf{17}, 
116 (2020).

\bibitem{ZitterbewegungJPhysConf}
A. J. Silenko, Zitterbewegung in quantum mechanics of Proca particles,
J. Phys.: Conf. Ser. \textbf{1435}, 012057 (2020).

\bibitem{ZitterbewegungBD}
J. D. Bjorken and S. D. Drell, \emph{Relativistic Quantum Mechanics}
(McGraw-Hill, New York, 1964).

\bibitem{Messiah}
A. Messiah, \emph{Quantum Mechanics}, Vol. II (Wiley, New York, 1958), Chap. XX, Sec. 37.

\bibitem{Thaller}
B. Thaller, \emph{The Dirac Equation} (Springer, Berlin, 1992).

\bibitem{ZitterbewegungKrekora}
P. Krekora, Q. Su, and R. Grobe, Relativistic Electron Localization and the Lack of Zitterbewegung, Phys. Rev. Lett. \textbf{93}, 043004 (2004).

\bibitem{Zitterbewegungbook}
R. F. O'Connell, Rotation and spin in physics, in \emph{General Relativity and John Archibald Wheeler}, edited by I. Ciufolini and R. Matzner (Springer, Berlin, 2010).

\bibitem{ZitterbewegungOC}
R. F. O'Connell, Zitterbewegung is not an observable, Mod. Phys. Lett. A \textbf{26}, 469 (2011).

\bibitem{PRAFW}
Liping Zou, Pengming Zhang, and A. J. Silenko, Position and spin in relativistic quantum mechanics, Phys. Rev. A \textbf{101}, 032117 (2020).

\bibitem{ZitterbewegungDeriglazov}
A. A. Deriglazov, Semiclassical description of relativistic spin without use of
Grassmann variables and the Dirac equation, arXiv:1107.0273; Spinning-particle model for the Dirac equation and the relativistic Zitterbewegung, Phys. Lett. A \textbf{376}, 309 (2012). 

\bibitem{ZitterbewegungKobakhidze}
A. Kobakhidze, A. Manning, A. Tureanu, Observable Zitterbewegung in curved spacetimes, Phys. Lett. B \textbf{757}, 84 (2016). 

\bibitem{NewtonWigner}
T. D. Newton and E. P. Wigner, Localized States for Elementary Systems, Rev. Mod. Phys. \textbf{21}, 400 (1949).

\bibitem{Wightman}
A. S. Wightman, On the Localizability of Quantum
Mechanical Systems, Rev. Mod. Phys. \textbf{34}, 845 (1962).

\bibitem{Unal}
N. \"{U}nal, A Simple Model of the Classical Zitterbewegung:
Photon Wave Function, Found. Phys. \textbf{27}, 
731 (1997).

\bibitem{Kobe}
D. H. Kobe, Zitterbewegung of a photon, Phys. Lett. A \textbf{253}, 7 (1999). 

\bibitem{ZhiYongWang1}
Zhi-Yong Wang and Cai-Dong Xiong, Zitterbewegung in quantum field theory, Chinese Phys. B \textbf{17}, 4170 (2008).

\bibitem{ZhiYongWang2}
Zhi-Yong Wang, Cai-Dong Xiong, and Qi Qiu, Photon wave function and Zitterbewegung, Phys. Rew. A \textbf{80}, 032118 (2009). 

\bibitem{ZhiYongWang3}
Zhi-Yong Wang, Cai-Dong Xiong, and Qi Qiu, Photonic zitterbewegung and its interpretation, Chinese Phys. B \textbf{21}, 020302 (2012).

\bibitem{BLP}
V. B. Berestetskii, E. M. Lifshitz, and L. P. Pitayevskii,
{\em Quantum Electrodynamics}, 2nd ed. (Pergamon, Oxford, 1982).

\bibitem{FV}
H. Feshbach and F. Villars, Elementary Relativistic Quantum Mechanics of Spin 0 and Spin 1/2 Particles, Rev. Mod. Phys. \textbf{30}, 24 (1958).

\bibitem{TMP2008}
A. J. Silenko, Hamilton operator and the semiclassical limit for
scalar particles in an electromagnetic field, Theor.
Math. Phys. \textbf{156}, 1308 
 [Teor. Mat. Fiz. \textbf{156}, 398 (2008)].

\bibitem{Honnefscalar}
A.\,J. Silenko, Scalar particle in general inertial and gravitational
fields and conformal invariance revisited,
Phys. Rev. D \textbf{88}, 045004 (2013).

\bibitem{Honnefscalarnew}
A.\,J. Silenko, New symmetry properties of pointlike scalar and Dirac particles, Phys. Rev. D \textbf{91}, 065012 (2015).

\bibitem{HonnefscalarLT}
A.\,J. Silenko, Quantum-Mechanical Description of Lense-Thirring Effect for Relativistic Scalar Particles, Phys. Part. Nucl. Lett.
\textbf{10}, 637 (2013).

\bibitem{SaTa}
M. Taketani and S. Sakata, On the Wave Equation of Meson, Proc. Phys. Math. Soc. Japan {\bf 22},
757 (1940).

\bibitem{SpinunitEDM}
A.\,J. Silenko, Quantum-mechanical description of spin-1 particles with electric dipole moments, Phys. Rev. D \textbf{87}, 073015 (2013).

\bibitem{Simulik} V.\,M. Simulik, I.\,Yu. Krivsky, Bosonic symmetries of the massless Dirac equation,
Adv. Appl. Clifford Alg. \textbf{8}, 
69 (1998).

\bibitem{Simulik1}
V.\,M. Simulik, I.\,Yu. Krivsky, On the extended real Clifford-Dirac algebra and new physically meaningful symmetries of the Dirac equations with nonzero mass, Reports of the National Academy of Sciences of Ukraine, No. 5, 82 (2010).

\bibitem{Simulik2}
I.\,Yu. Krivsky, V.\,M. Simulik, Fermi-Bose duality of the Dirac equation and extended
real Clifford-Dirac algebra, Condensed Matter Physics \textbf{13}, 43101 (2010).

\bibitem{Simulik3} V.\,M. Simulik, I.\,Yu. Krivsky, Bosonic symmetries of the Dirac equation,
Phys. Lett. A \textbf{375}, 
2479 (2011).

\bibitem{Simulik4}
V.\,M. Simulik, I.\,Yu. Krivsky, I.\,L. Lamer, Bosonic symmetries, solutions, and conservation laws for the Dirac equation with nonzero mass,
Ukrainian Journal of Physics 
\textbf{58}, 523 (2013).

\bibitem{Simulik5}
V.\,M. Simulik, I.\,Yu. Krivsky, I.\,L. Lamer, Application of the generalized Clifford-Dirac algebra to the proof of the Dirac equation Fermi-Bose duality,
TWMS J. Appl. Eng. Math. 
\textbf{3}, 46 (2013).

\bibitem{Simuliknew}
V.\,M. Simulik, I.\,Yu. Krivsky, I.\,L. Lamer, Some statistical aspects of the spinor field Fermi-Bose duality, Condensed Matter Physics \textbf{15}, 43101 (2012).

\bibitem{OConnell}
R. F. O'Connell, E. P. Wigner, On the relation between momentum and velocity for elementary systems, Phys. Lett. A  \textbf{61}, 353 (1977); Position operators for systems exhibiting the special relativistic relation between momentum and velocity, Phys. Lett. A  \textbf{67}, 319 (1978). 	

\bibitem{FW}
L. L. Foldy, S. A. Wouthuysen, On the Dirac Theory of Spin 1/2
Particles and Its Non-Relativistic Limit, Phys. Rev.
\textbf{78}, 29 (1950).

\bibitem{JMP}
A.\,J. Silenko, Foldy-Wouthuysen transformation for
relativistic particles in external fields, J. Math. Phys. {\bf 44}, 2952 (2003).

\bibitem{JINRLett12}
A. J. Silenko, Classical limit of relativistic quantum mechanical equations in the Foldy-Wouthuysen representation, Pis'ma Zh. Fiz. Elem. Chast. Atom. Yadra \textbf{10},
144 (2013) [Phys. Part. Nucl. Lett. \textbf{10}, 91 (2013)].

\bibitem{ExpectationValue} A. J. Silenko, Energy expectation values of a 
particle in nonstationary fields, Phys. Rev. A \textbf{91}, 012111 (2015).

\bibitem{relativisticFW}
A. J. Silenko, General properties of the Foldy-Wouthuysen transformation and
applicability of the corrected original Foldy-Wouthuysen method, Phys. Rev. A \textbf{93}, 022108 (2016).

\bibitem{LandauPeierls}
L. Landau and R. Peierls, Quantenelektrodynamik im Konfigurationsraum, Z. Physik \textbf{62}, 188 (1930);
Erweiterung des Unbestimmtheitsprinzips f\"{u}r die relativistische Quantentheorie. Z. Physik \textbf{69}, 56 (1931). 

\bibitem{Ali}
S. T. Ali, Stochastic localization, quantum mechanics on phase space and quantum space-time, Riv. Nuovo Cim. \textbf{8}, 1 (1985).

\bibitem{Ingall}
J. E. M. lngall, The Newton-Wigner and Wightman Localization
of the Photon, Found. Phys. \textbf{26}, 1003 (1996).

\bibitem{footnote}
F. Pampaloni, J. Enderlein, Gaussian, Hermite-Gaussian, and
Laguerre-Gaussian beams: A primer, arXiv:physics/0410021 (2004); K. Y. Bliokh, F. Nori, Transverse and longitudinal angular momenta of light, Phys. Rep. \textbf{592}, 1 (2015). 

\bibitem{PRA2015}
A.\,J. Silenko, General method of the relativistic Foldy-Wouthuysen transformation
and proof of validity of the Foldy-Wouthuysen Hamiltonian, Phys. Rev. A \textbf{91}, 022103 (2015).

\bibitem{Allen}
L. Allen, M. W. Beijersbergen, R. J. C. Spreeuw, J. P. Woerdman,
Orbital angular momentum of light and the transformation
of Laguerre-Gaussian laser modes, Phys. Rev. A \textbf{45}, 8185
(1992).

\bibitem{Siegman}
A. E. Siegman, \emph{Lasers} (University Science Books,
Sausalito, 1986).

\bibitem{BialBirOpt}
I. Bialynicki-Birula, Z. Bialynicka-Birula, Quantum-mechanical description of optical beams,
J. Opt. \textbf{19}, 125201 (2017).

\bibitem{Loudon}
R. Loudon, \emph{The Quantum Theory of Light}, 2rd ed. (Oxford Univ. Press, Oxford, 2000).

\bibitem{Srednicki}
M. Srednicki, \emph{Quantum Field Theory} (Cambridge Univ. Press, Cambridge, 2007).

\bibitem{BialynickiBirulaPhysScr}
I. Bialynicki-Birula, Z. Bialynicka-Birula,
Quantum numbers and spectra of structured light,
Phys. Scr. \textbf{93}, 104005 (2018).

\bibitem{Dirac-like}
 J. R. Oppenheimer, Note on light quanta and the electromagnetic field, Phys. Rev. \textbf{38}, 725 (1931); G. Moli\`{e}re,  Laufende elektromagnetische Multipolwellen und eine neue methods der Feld-Quantisierung, Ann. Phys. (Lpz.) \textbf{6}, 146 (1949); W. J. Archibald, Field equations from particle equations, Can. J. Phys. \textbf{33}, 565 (1955); I. Bialynicki-Birula, On the wave function of the photon, Acta Phys. Polon. A \textbf{86}, 97 (1994); The photon wave function, in \emph{Coherence and Quantum Optics VII}, edited by J. H. Eberly, L. Mandel, and E. Wolf (Plenum, New York, 1996), p. 313; Photon wave function, Prog. Opt. \textbf{36}, 245 (1996).

\bibitem{Barnett-FWQM}
S. M. Barnett, Optical Dirac equation, New J. Phys. \textbf{16},
093008 (2014).

\bibitem{YB}
J. A. Young and S. A. Bludman, Electromagnetic Properties of a Charged Vector
Meson, Phys. Rev. {\bf 131}, 2326 (1963).

\bibitem{spin1}
A. J. Silenko, Polarization of spin-1 particles in a uniform magnetic field,
Eur. Phys. J. C \textbf{57}, 595 (2008); High precision description and new properties of a spin-1 particle in a magnetic field, Phys. Rev. D \textbf{89},
121701(R) (2014); Relativistic quantum mechanics of a Proca particle
in Riemannian spacetimes, Phys. Rev. D \textbf{98},
025014 (2018).

\bibitem{Anderson}
P. W. Anderson, Local Moments and Localized States, Science \textbf{201}, 307 (1978).

\bibitem{Born-Wolf}
M. Born and E. Wolf, \emph{Principles of Optics: Electromagnetic Theory of Propagation,
Interference and Diffraction of Light}, 7th ed. (Cambridge University Press, Cambridge,
1999), Chap. 1.

\bibitem{Bouchard}
F. Bouchard, J. Harris, H. Mand, R. W. Boyd, and E. Karimi,
Observation of subluminal twisted light in
vacuum, Optica \textbf{3}, 351 (2016).

\bibitem{Chen}
Z. Chen, Y. K. Ho, P. X. Wang, Q. Kong, Y. J. Xie, W. Wang,
and J. J. Xu, A formula on phase velocity of waves and application,
Appl. Phys. Lett. \textbf{88}, 121125 (2006).

\bibitem{HuangWuHu}
S. Huang, F. Wu, and B. Hu, Formula for the phase velocity of electromagnetic waves,
Phys. Rev. E \textbf{79}, 047601 (2009).

\bibitem{Saari}
P. Saari, Reexamination of group velocities of structured light pulses,
Phys. Rev. A \textbf{97}, 063824 (2018).

\bibitem{WangScheidHo}
P. X. Wang, W. Scheid, and Y. K. Ho,
Electron capture acceleration channel in a slit laser beam,
Appl. Phys. Lett. \textbf{90}, 111113 (2007).

\bibitem{ZhangWangXie}
X. P. Zhang, W. Wang, Y. J. Xie, P. X. Wang, Q. Kong, Y. K. Ho,
Field properties and vacuum electron acceleration in a laser beam of high-order Laguerre-Gaussian mode,
Opt. Commun. \textbf{281}, 4103 (2008).

\bibitem{WangHoYuan}
P. X. Wang, Y. K. Ho, X. Q. Yuan, Q. Kong, N. Cao, A. M. Sessler, E. Esarey, and Y. Nishida,
Vacuum electron acceleration by an intense laser,
Appl. Phys. Lett. \textbf{78}, 2253 (2001).

\bibitem{WangQian}
K. Wang, L. Qian, P. Qiu, and H. Zhu,
Phase-velocity measurement of a tightly focused Gaussian beam by use of sum
frequency generation,
Appl. Phys. Lett. \textbf{92}, 121114 (2008).

\bibitem{LightArXiv}
A. J. Silenko, Pengming Zhang, and Liping Zou, Relativistic quantum-mechanical description of twisted paraxial electron and photon beams, Phys. Rev. A \textbf{100}, 030101(R) (2019).

\bibitem{Hejtler}
W. Hejtler, \emph{The quantum theory of radiation}, 3rd ed. (Clarendon Press, Oxford, 1954). 

\bibitem{DarwinC}
C. G. Darwin, Notes on the theory of radiation, Proc. R. Soc. A \textbf{136}, 36 (1932). 

\bibitem{MohrP}
P. J. Mohr, Solutions of the Maxwell equations and photon wave functions, Ann. Phys. (N.Y.) \textbf{325}, 607 (2010). 

\bibitem{Weinberg}
S. Weinberg, Feynman Rules for Any Spin, Phys. Rev. \textbf{133}, B1318 (1964).

\bibitem{JINRLDFW}
A.\,J. Silenko, Connection between wave functions in the Dirac and Foldy-Wouthuysen representations, Phys. Part. Nuclei Lett. \textbf{5}, 501 (2008). 

\bibitem{Sachs}
M. Sachs, The Quantum Negative Energy Problem Revisited,  Ann. Fond. Louis de Broglie, \textbf{30}, 381 (2005). 

\bibitem{Feynman}
R. P. Feynman, \emph{The quantum theory of radiation} (Taylor and Francis Group, Boka Raton, 2018). 

\bibitem{Debergh}
N. Debergh, J.-P. Petit and G. D?Agostini, On evidence for negative energies and masses in the Dirac equation through a unitary time-reversal operator, J. Phys. Commun. \textbf{2}, 115012 (2018).

\bibitem{Wolf}
G. Wolf, Review of High Energy Diffraction in Real and Virtual Photon Proton scattering at HERA, Rept. Prog. Phys. \textbf{73}, 116202 (2010).

\bibitem{Mart}
T. Mart, Electromagnetic Productions of the Hyperon and the Hypertriton Using Real and Virtual Photons, EPJ Web Conf., \textbf{3}, 07002 (2010). 

\bibitem{Liu}
Wenjian Liu, Perspectives of relativistic quantum chemistry: the negative energy cat smiles, Physical Chemistry Chemical Physics \textbf{14}, 35 (2012). 

\bibitem{Liberato}
S. De Liberato, Virtual photons in the ground state of a dissipative system, Nature Comm. \textbf{8}, 1465 (2017).

\bibitem{Alekseev}
V. I. Alekseev, A. N. Eliseyev, E. Irribarra, I. A. Kishina, A. S. Klyueva, A. S. Kubankin, R. M. Nazhmudinov, and S. V. Trofymenko, Diffraction of virtual and real photons, J. Instrum. \textbf{15}, C03009 (2020).

\bibitem{ZitterbewegungRo}
M. E. Rose, \emph{Relativistic Electron Theory} (Wiley, New York,
1961).

\bibitem{ZitterbewegungCapelle}
K. Capelle, Relativistic fluctuations and anomalous Darwin terms in superconductors, Phys. Rev. B \textbf{63}, 052503 (2001).

\bibitem{PRA2008}
A.\,J. Silenko, Foldy-Wouthyusen transformation and semiclassical limit for relativistic particles
in strong external fields, Phys. Rev. A \textbf{77}, 012116 (2008).


\bibitem{RPJ}
A.\,J. Silenko, Quantum-mechanical description of the
electromagnetic interaction of relativistic particles with
electric and magnetic dipole moments, Russ. Phys. J. \textbf{48},
788 (2005). 

\bibitem{PhysRevDspinunit}
A. J. Silenko, Quantum-mechanical description of spin-1 particles with electric dipole moments, Phys. Rev. D \textbf{87}, 073015 (2013).

\bibitem{JETP}
A. J. Silenko, The motion of particle spin in a nonuniform electromagnetic field, JETP \textbf{96},
775 (2003).

\bibitem{classEDM}
D. F. Nelson, A. A. Schupp, R. W. Pidd and H. R. Crane,
Search for an Electric Dipole Moment of the Electron.
Phys. Rev. Lett. {\bf 2}, 492 (1959);
T. Fukuyama and A. J. Silenko, Derivation of Generalized
Thomas-Bargmann-Michel-Telegdi Equation for a Particle with Electric
Dipole Moment, Int. J. Mod. Phys. A \textbf{28}, 1350147 (2013);
A. J. Silenko, Spin precession of a particle with an electric dipole moment:
contributions from classical electrodynamics and from the Thomas effect,
Phys. Scripta {\bf 90}, 065303 (2015).

\bibitem{Pryce}
M. H. L. Pryce, The mass-centre in the restricted theory of relativity and its connexion with the quantum theory of elementary particles,  Proc. R. Soc. London A \textbf{195}, 62 (1948).

\bibitem{SuttorpDeGroot} L. G. Suttorp and S. R. De Groot, Covariant equations of motion for a charged particle with a magnetic dipole moment, Nuovo Cimento A \textbf{65},
245 (1970).

\bibitem{DeKerfBauerle} E. A. De Kerf and G. G. A. B\"{a}uerle, A position operator for a relativistic particle with spin, Physica (Amsterdam)
\textbf{57}, 121 (1972).

\bibitem{KhPom} I. B. Khriplovich and A. A. Pomeransky, Equations of
motion of spinning relativistic particle in external fields,
Phys. Lett. A \textbf{216}, 7 (1996).

\bibitem{PomKh} A. A. Pomeransky and I. B. Khriplovich, Equations of
motion of spinning relativistic particle in external fields,
Zh. Eksp. Teor. Fiz. \textbf{113}, 1537 (1998) [J. Exp. Theor. Phys.
\textbf{86}, 839 (1998)].

\bibitem{Obzor}
A. A.~Pomeransky, R. A.~Senkov, and I. B.~Khrip\-lo\-vich,
Spinning relativistic particles in external fields,
Usp. Fiz. Nauk \textbf{43}, 1129 (2000) 
 [Phys. Usp. \textbf{43}, 1055 (2000)]. 

\end{thebibliography}
\end{document}